\documentclass[fleqn,usenatbib]{mnras}

\usepackage{newtxtext,newtxmath}
\usepackage{makecell} % 表格中一格内容分为多行书写
\usepackage{marginnote}
\usepackage[T1]{fontenc}

\DeclareRobustCommand{\VAN}[3]{#2}
\let\VANthebibliography\thebibliography
\def\thebibliography{\DeclareRobustCommand{\VAN}[3]{##3}\VANthebibliography}

\usepackage{graphicx}	% Including figure files
\usepackage{amsmath}	% Advanced maths commands
\usepackage{threeparttable}
\newcommand{\fermi}{\textit{Fermi}-{\rm LAT}}

\newcommand{\gray}{$\gamma$-ray}
\newcommand{\grays}{$\gamma$-rays}
\newcommand{\xray}{$\rm X$-ray}

\newcommand{\msun}{\mbox{$M_\odot$}}

\def\deg{\hbox{$^\circ$}}

\title[GeV gamma-ray towards Vela Jr]{GeV \gray\ emission in the field of the shell-type supernova remnant Vela Jr revisited}
\author[Ge et. al]{Ting-Ting Ge$^{1,}$$^{2}$,
Qi-Hang Wu$^{3}$,
Pak-Hin Thomas Tam$^{1,}$$^{2}$\thanks{E-mail:tanbxuan@mail.sysu.edu.cn} \& Jie Feng$^{4}$\thanks{E-mail:fengj77@mail.sysu.edu.cn},
Hai-Feng Zhou$^{5}$,
Kai Wang$^{6}$,
\newauthor{Su-Jie Lin}$^{1,}$$^{2}$
\\
$^{1}$School of Physics and Astronomy, Sun Yat-sen University, Zhuhai 519082, China\\
$^{2}$CSST Science centre for the Guangdong-Hong Kong-Macau Greater Bay Area, Sun Yat-Sen University, Zhuhai 519082, China\\
$^{3}$Department of Astronomy, Yunnan University, and Key Laboratory of Astroparticle Physics of Yunnan Province, Kunming, 650091, People's Republic of China\\
$^{4}$School of Science, Shenzhen Campus of Sun Yat-sen University, Shenzhen 518107, China\\
$^{5}$School of Physics, Sun Yat-Sen University, Guangzhou 510275, China\\
$^{6}$School of Astronomy and Space Science, Nanjing University, Nanjing 210023, Jiangsu, China
}
\pubyear{2025}

\begin{document}
\label{firstpage}
\pagerange{\pageref{firstpage}--\pageref{lastpage}}
\maketitle{}

\begin{abstract}
We present an updated analysis of the gigaelectronvolt (GeV) \gray\ emission from the shell-type supernova remnant (SNR) RX~J0852.0-4622 (Vela Jr) using 15 yr of Fermi Large Area Telescope (\fermi) data. 
We quantitatively model the GeV morphology and find that it is best described by the masked H.E.S.S. shell template, indicating that the embedded pulsar wind nebula (PWN) contributes little to the GeV flux.
The 0.1--500\,GeV spectrum is well fitted by a hard power law with a photon index of $1.77 \pm 0.03$ and connects smoothly to the teraelectronvolt (TeV) spectrum, confirming previous results with improved precision.
We further construct an independent eROSITA shell template and derive the 1--5\,keV X-ray spectral energy distribution (SED) of the whole remnant, which provides new constraints on the synchrotron emission.
We model the multi-wavelength (MWL) SED with a pure leptonic model and a hybrid lepton-hadron model. While the pure leptonic model reproduces the overall broadband shape, the hybrid model provides a better statistical description of the same dataset, supporting a mixed-origin picture in which the hadronic contribution is mainly relevant in the GeV band and the TeV emission remains predominantly leptonic. 

\end{abstract}


%%%%%%%%%%%%%%%%%%%%%%%%%%%%%%%%%%%%%%%%%%%%%%%%%%

%%%%%%%%%%%%%%%%% BODY OF PAPER %%%%%%%%%%%%%%%%%%

%%%%%%%%%%%%%%%%% introduction %%%%%%%%%%%%%%%%%%
\section{Introduction}
Supernova remnants (SNRs) exhibit shock fronts that appear as shells comprised of arc-like structures, and their forward shocks are widely considered prime sites of diffusive shock acceleration, an efficient mechanism for accelerating charged particles to relativistic energies \citep{1977DSA1, 1977DSA2, 1978DSA3, 1978DSA4, 1987Blandford, 2015Hewitt}.
In the past decade, measurements of a handful of shell-type SNRs in very high energy gamma rays have provided unique insights into the acceleration process.
Observationally, several SNRs have been detected with a clear shell type morphology resolved in GeV-TeV \grays\ energies, that is, RCW 86 \citep{2018HESS_RCW86}, RX J1713.7-3946 \citep{2018HESS_RXJ1713}, RX J0852.0-4622 \citep{2011Tanaka,2018HESS}, SN 1006 \citep{2016Xing}, and HESS J1731-347 \citep{2017Condon}.
Meanwhile, SNR candidates have been proposed purely based on the shell-type appearance at TeV energies \citep{2018HESS_612_A8}.

The SNR RX J0852.0-4622, also referred to as G266.2-1.2 or Vela Jr, overlaps the southeast corner of the Vela SNR \citep{2023Mayer} and has been suggested to be a core-collapse SNR \citep{1998Aschenbach}.
It is classified to be a young shell-type SNR discovered on the Galactic plane in the ROentgen SATellite (ROSAT) All-Sky Survey \citep{1998Aschenbach,1996Pfeffermann}.
The distance to the remnant and its age are still debated in the literature, but the range of possible values is narrowing \citep{2008Katsuda,2015Allen}. 
Based on the \textit{XMM-Newton} data and assuming a shock velocity of 3000 km s$^{-1}$, \citet{2008Katsuda} estimate the age and distance of the shell to be 1.7-4.3 kyr and $\sim750$ pc, respectively.
By analyzing \textit{Chandra} data, \citet{2015Allen} derived an age of 2.4-5.1 kyr and the distance of 0.5-1 kpc, which are in agreement with the study by \citet{2008Katsuda}. 
Using XMM-Newton and eROSITA observations, \citet{2023Camilloni} derived an age of 2.4-5.1 kyr and a distance of $\sim1.1$ kpc.
Recently, \citet{2025Suherli} derived a more precise distance constraint of $1.41 \pm 0.14$ kpc by linking Vela Jr to the Gaia-based distance of the associated Herbig-Haro source Ve 7-27. We adopt 1.41 kpc as the reference distance when converting to physical quantities throughout this work.

\begin{table*}
   %\centering
   \caption{Summary of previous high-energy studies of the SNR RX J0852.0-4622.}
   \begin{tabular}{ccccc}
        \hline\hline
        & Instrument & Age (kyr) & Distance (kpc) & References \\ \hline
        Optical & VLT/MUSE + Gaia & $\sim$1.6--3.3 & $1.41\pm0.14$ & \citet{2025Suherli} \\
        X-ray & ROSAT & 0.5-1.1 & 0.08-0.5 & \makecell[c]{\citet{1998Aschenbach},\\ \citet{1999Aschenbach}} \\
        &&&&\\
        & ASCA & 0.63-0.97 & 1-2 & \makecell[c]{\citet{2000Tsunemi,2001Slane}\\ \citet{2013Lee}} \\
        % &&&&\\
        % & BeppoSAX &  & northermal &  \citet{2001Mereghetti}\\
        &&&&\\
        & Chandra & 2.4-5.1 & 0.5-1 & \makecell[c]{\citet{2001Pavlov,2002Kargaltsev},\\ \citet{2005Bamba,2005Iyudin},\\ \citet{2010Pannuti,2013Lee},\\ \citet{2015Allen}} \\
        &&&&\\
        & XMM-Newton & 1.7-4.3 & 0.75 & \makecell[c]{\citet{2008Katsuda,2013Acero},\\ \citet{2013Kishishita,2023Camilloni}} \\
        &&&&\\
        & eROSITA & 2.4-5.1 & 1.1 & \citet{2023Camilloni}\\
        MeV & COMPTEL & $\sim0.68$ & $\sim0.2$ & \citet{1998Iyudin} \\
        &&&&\\  
        %\cdashline{1-5}
        \hline
        & Instrument & Morphology & Spectrum (index) & References \\ \hline
        X-ray & Suzaku & shell-like & PL ($2.92\pm0.01$) & \makecell[c]{\citet{2016Takeda,2017Fukui},\\ \citet{2024Fukui}} \\
        GeV & Fermi-LAT & HESS template & PL ($1.85\pm0.20$) & \citet{2011Tanaka} \\
        &  & disk (0.98$^\circ$) & PL ($1.83\pm0.08$) & \citet{2017Ackermann} \\
        TeV & CANGAROO-II & NW$^a$ & PL (4.3$^{+4.4}_{-1.7}$) & \citet{2005Katagiri}\\
        &  & shell-like & PL ($2.2\pm0.42$) & \citet{2006Enomoto}\\
        % & HESS & shell-like & PL (2.1-2.24) & \makecell[c]{\citet{2005Aharonian,2007Aharonian}, \\ \citet{2018HESS}}\\
        % & HESS & shell-like & PL ($2.1\pm0.22-2.24\pm0.16$) & \citet{2005Aharonian,2007Aharonian} \\
        & HESS & shell-like & PL ($2.24\pm0.16$) & \citet{2005Aharonian,2007Aharonian} \\
        &  & shell-like & PLEC ($1.81\pm0.08$)$^b$ & \citet{2018HESS} \\
        % &  & HESS template & 52\%:48\%$^c$ & \citet{2017Fukui,2024Fukui}\\
        \hline\hline
   \end{tabular}
   \label{tab:known info}
       \begin{tablenotes}
           \item $^a$ The data only cover the northwestern region of the SNR.
           \item $^b$ The energy cut-off is $6.7\pm1.2$ TeV.
           % \item $^c$ Denotes the ratio of hadronic to leptonic contribution to the TeV \grays.
       \end{tablenotes}
\end{table*}

An energetic rotation-powered pulsar PSR J0855-4644 ($\dot{E} = 1.1 \times 10^{36}\ \rm erg\ s^{-1}$) lies on the southeast rim of the RX J0852.0-4622 \citep{2003Kramer}.
\xray\ observations with \textit{XMM-Newton} have revealed a PWN with an extension of $150\arcsec$ surrounding the PSR J0855-4644 \citep{2013Acero}.
In addition, a candidate neutron star CXOU J085201.4-461753 is located near the centre of the SNR RX J0852.0-4622 \citep{2001Pavlov}.
A number of X-ray surveys (as shown in Table~\ref{tab:known info}) have suggested different compact objects, such as the PSR J0855-4644 and the candidate neutron star CXOU J085201.4-461753, which may be associated with the SNR \citep{2001Pavlov, 2002Kargaltsev}.
As of yet, no firm association has been established between the two compact objects and the remnant, despite their compatible distances \citep{2013Acero, 2001Pavlov, 2002Kargaltse}.
In light of the new X-ray data, \citet{2023Camilloni} have refined the remnant's geometrical centre that directly affects the measured proper motions.
Therefore, if CXOU J085201.4-461753 is associated with the SNR, there was no significant displacement from its birthplace.

%%%%
In the \gray\ energy range, RX J0852.0-4622 exhibits a shell-like TeV \gray\ distribution \citep{2018HESS}, which is similar to the \xray\ shell, analogous to another shell-type SNR, RX J1713.7-3946 \citep{2018HESS_RXJ1713}.
A detailed analysis of the H.E.S.S. data at and around the remnant's vicinity \citep{2018HESS} shows that the TeV \gray\ spectrum can smoothly connect to the hard spectrum in the GeV band detected by \fermi\ \citep{2011Tanaka}.
Concurrently, the absence of thermal \xray\ emission supports the leptonic origin hypothesis as well \citep{2001Slane}.

However, the origin of \gray\ emission from the SNR remains obscure.
\citet{2017Fukui} analyzed the interstellar medium (ISM) proton distribution toward RX J0852.0-4622, which includes both molecular and atomic gas, and showed that the total ISM protons associated with the SNR correspond well spatially to the \gray\ distribution.
In addition, radio-continuum observations from RX J0852.0-4622 show spatial concordance with the \xray\ and \gray\ emissions reported in \citet{2018Maxted}.  At the northwestern edge of RX J0852.0-4622, the radio spectral index becomes progressively flatter as it approaches a neighboring molecular clump, which is conjectured to be related to RX J0852.0-4622 as well \citep{2018Maxted}. 
This flattening may indicate a shock-cloud interaction and enhanced target density for pp collisions, potentially increasing a hadronic $\gamma$-ray contribution; however, it is not a unique discriminator between hadronic and leptonic scenarios, and MWL evidence for such an interaction remains inconclusive \citep{2018Maxted}.
%This provides strong evidence for a significant hadronic \gray\ component in the \gray\ emission from RX J0852.0-4622.
Since the SNR RX J0852.0-4622 is an extensively studied object, particularly in the very high energy domain, that is, GeV and TeV \grays, we have compiled relevant \gray\ research for the SNR. The results are summarized in Table~\ref{tab:known info}.

Overall, both hadronic (proton-proton interactions with subsequent $\pi_{0}$ decay) and leptonic (inverse Compton (IC) scattering of relativistic electrons on ambient radiation fields) scenarios  have been reported to explain the \gray\ emission from RX J0852.0-4622 \citep{2007Aharonian,2011Tanaka,2018HESS}.
Recently, \citet{2024Fukui} modelled the TeV $\gamma$-rays counts as a linear combination of hadronic $\gamma$-rays traced by the interstellar gas column density and leptonic $\gamma$-rays traced by the Suzaku non-thermal X-ray emission, under these assumptions they inferred an approximately equal hadronic and leptonic contribution $\sim 5:5$.
\fermi\ can provide important information on GeV \gray\ emission, which is crucial to constrain the possible emission models.

In this paper, we report the analysis results of the GeV \gray\ emission toward the remnant RX J0852.0-4622 using more than 15 years of \fermi\ data, as demonstrated in sect.\ref{sec:data}. In sect.\ref{sec:origin}, we investigate the possible origin of the \gray\ emission at the GeV-TeV energy range. Finally, we discuss how our study adds to our knowledge of this remnant at present and summarize the main conclusions in sect.\ref{sec:Disc}.

%----------------------------------------------------- FIGURE 1
\begin{figure*}
%\centering
\includegraphics[scale=0.35]{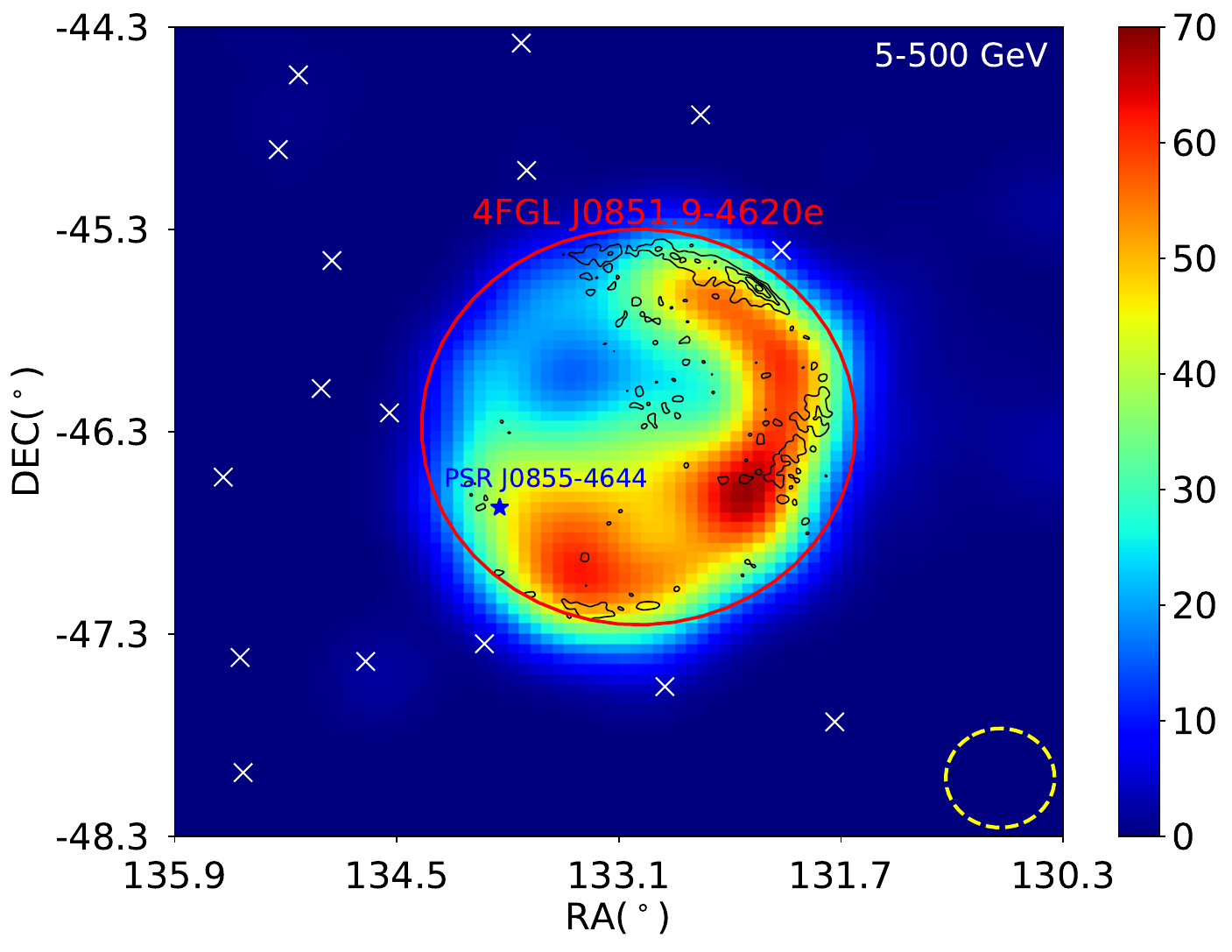}
\includegraphics[scale=0.35]{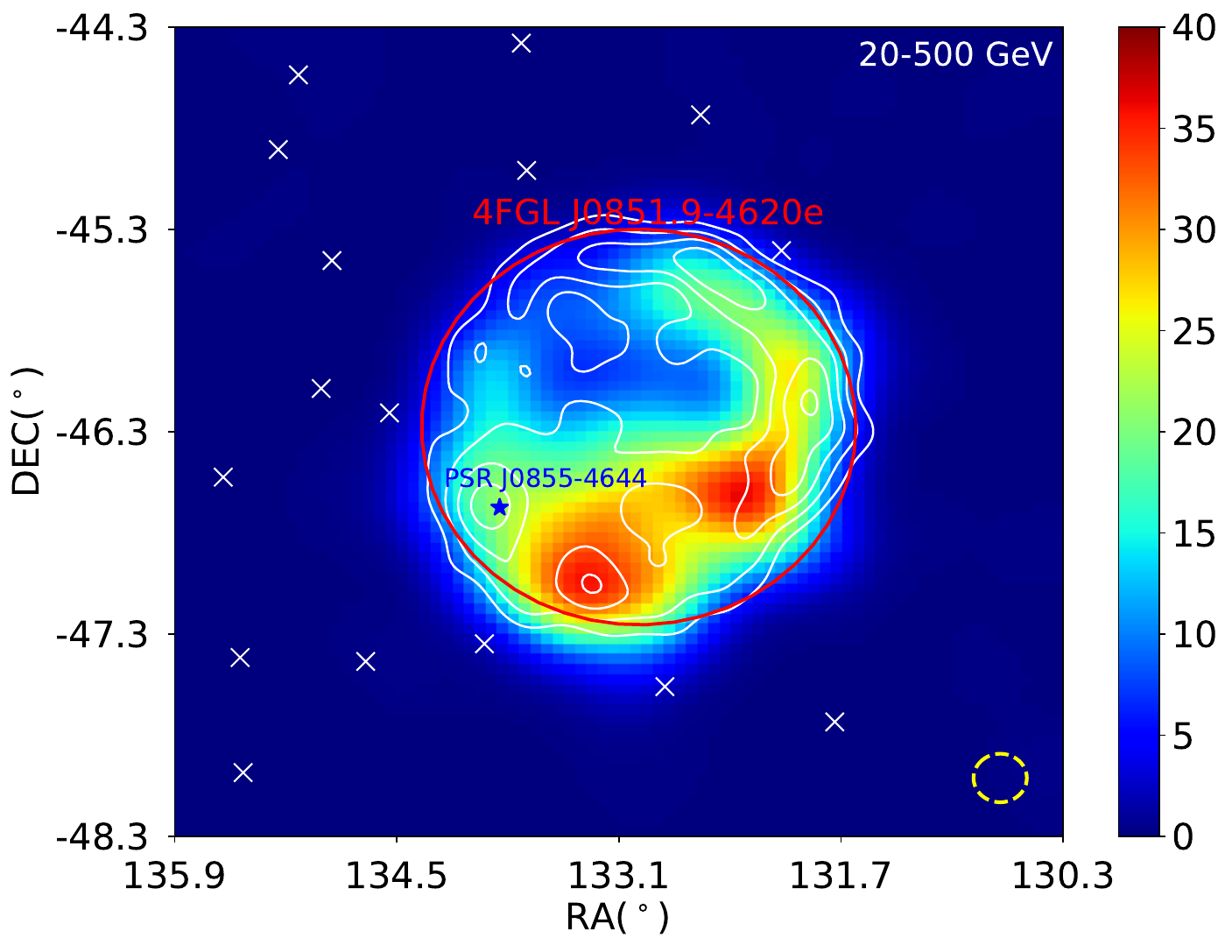}
\caption{\fermi\ TS maps of RX J0852.0-4622 in the energy range of 5--500\,GeV (left panel) and 20--500\,GeV (right panel). The size is that of a $4\deg \times 4\deg$ region smoothed with a Gaussian filter of $1\deg$, and each pixel is $0.05\deg \times 0.05\deg$ in size. All white crosses represent the 4FGL-DR4 sources within the region. The red circle represents the extended source 4FGL J0851.9-4620e related to RX J0852.0-4622. The blue star indicates the position of PSR J0855-4644. In the left panel, X-ray contours from the first eROSITA All-Sky Survey data in the 1--8\,keV energy range are shown in black. In the right panel, the shape of the SNR shell observed by H.E.S.S. \citep{2018HESS} is depicted as white significance contours at every 2$\sigma$ from 3$\sigma$ as the lowest. The yellow dashed circle in the right corner of both panels illustrates the PSF size of the instrument at the lower energy cut used in the analysis.}
\label{fig:cmap}
\end{figure*}
%%%%%%%%%%%%%%%%% data reduction and analysis %%%%%%%%%%%%%%%%%%
\section{\fermi\ data analysis}
\label{sec:data}

\fermi\ has been continuously monitoring the sky since 2008 and is sensitive to \grays\ from 20 MeV to over 300 GeV \citep{2009Atwood}. We select the latest Pass 8 data \citep{2018Bruel} at and around the SNR RX J0852.0-4622 region from August 4, 2008 (MET 239557417) until May 8, 2023 (MET 705221903) and use the standard LAT analysis software package $\it v11r5p3$\footnote{\url{https://fermi.gsfc.nasa.gov/ssc/data/analysis/software/}}. The event class ``P8R3\_SOURCE'' (evclass=128) and event type FRONT + BACK (evtype=3) are used, with the standard data quality selection criteria $\rm (DATA\_QUAL > 0) \&\& (LAT\_CONFIG == 1)$. In order to reduce the \gray\ contamination from the Earth's albedo, only the events with zenith angles less than 90$\deg$ are included in the analysis. 
In this work, we use the Python module \footnote{\url{https://fermi.gsfc.nasa.gov/ssc/data/analysis/scitools/python_tutorial.html}} which implements a maximum likelihood optimization technique for a standard binned analysis. 
Data within a $14\deg \times 14\deg$ square region of interest (ROI), centered at the centroid of 4FGL J0851.9-4620e, adopted by Fermi collaboration to model RX J0852.0-4622 (a uniform disk with R.A.=133.08\deg, Dec=$-46.34\deg$, and radius 0.98\deg), are considered for our event subselection.
The exposure map of the entire sky is calculated with the instrument response functions (IRFs) ``P8R3\_SOURCE\_V3''. 
To estimate the \gray\ background, we include the recently released \fermi\ 14-year Source Catalog of point-like and extended sources (4FGL-DR4, \citet{2022Abdollahi, 2023Ballet}), the diffuse Galactic interstellar emission {$gll\_iem\_v07.fits$} and the isotropic extragalactic emission {$iso\_P8R3\_SOURCE\_V3\_v1.txt$} \footnote{\url{https://fermi.gsfc.nasa.gov/ssc/data/access/lat/BackgroundModels.html}}. 
All spectral parameters of the sources within $4\deg$ from the centre of the ROI, as well as normalizations of the Galactic and extragalactic background components, are set free during the fitting process.

%%%%。  Table 2   %%%%%
%\begin{table*}
%\centering
%\caption{Results of spatial analyses (5--500\,GeV) for different models.}
%\begin{tabular}{cccccccc}
%\hline
%Model &  & -$\log({\cal L})$ & $\rm TS$ & D.o.f.  & $\rm \Delta AIC$ \\
%\hline
           %& Fermibkg         & -79103 & - & 80 & - \\
%model 1    & 4FGL-DR4   ($R_{\rm disk} = 0.98^{\circ}$)      & -80208 & 2210 & 85   & 0 \\
%model 2    & uniform disk ($R_{\rm disk} = 1.02^{\circ}$)      & -80218 & 2230 & 85  & 20\\
%model 3    & Gaussian template ($R_{\rm sigma} = 0.7^{\circ}$)      & -80186 & 2166 & 85  &-44 \\
%model 4    & eROSITA template & -80237 & 2268 & 82  & 64 \\
%model 5    & H.E.S.S. template    & -80264 & 2322 & 82  & 118\\
%model 6  & masked H.E.S.S. template  & -80267 & 2326 & 82  & 124  \\
%\hline
%\end{tabular}
%    \begin{tablenotes}
%    \item D.o.f. denotes the number of degrees of freedom. 
%    \end{tablenotes}          
%\label{tab:likelihood}
%\end{table*}

\begin{table*}
\centering
\caption{Results of spatial analyses (5--500\,GeV) for different models.}
\begin{tabular}{cccccccc}
\hline
Model &   & $\rm TS$ & D.o.f.  & $\rm \Delta AIC$ & $\rm R.A.$, $\rm Dec$ & Extension\\
\hline
           %& Fermibkg         & -79103 & - & 80 & - \\
model 1    & 4FGL-DR4      & 2210 & 85   & 0 & 133.08\deg, $-46.34\deg$ & $R_{\rm disk} = 0.98^{\circ}$ \\
model 2    & uniform disk   & 2230 & 85  & 20 & $132.94\deg\pm 0.02\deg$, $-46.34\deg\pm 0.02\deg$ & $R_{\rm disk} = 1.02^{\circ}\pm 0.01 ^{\circ}$\\
model 3    & Gaussian template   & 2166 & 85  &$-44$ & $132.89\deg \pm 0.02\deg$, $-46.34\deg\pm0.05\deg$ & $R_{\rm sigma} = 0.70\deg\pm0.05\deg$  \\
model 4    & eROSITA template & 2268 & 82  & 64 & - & - \\
model 5    & H.E.S.S. template   & 2322 & 82  & 118 & - &-\\
model 6  & masked H.E.S.S. template  & 2326 & 82  & 124 & - & - \\
\hline
\end{tabular}
    \begin{tablenotes}
    \item D.o.f. denotes the number of degrees of freedom. 
    \end{tablenotes}          
\label{tab:likelihood}
\end{table*}

\subsection{Spatial analysis}

We use events in the 5--500\,GeV energy range to study the spatial distribution of the GeV \gray\ emission from the vicinity of RX J0852.0-4622. 
In the 4FGL-DR4 catalog, RX J0852.0-4622 is modelled as an extended \gray\ source, 4FGL J0851.9-4620e (hereafter model 1), which is listed as the GeV counterpart of the remnant. Previous \fermi\ studies identified RX J0852.0-4622 as a spatially extended GeV SNR with a morphology that broadly traces the shell seen at other wavelengths \citep{2011Tanaka, 2017Ackermann}.
We generate the TS maps using the gttsmap tool in the Fermitools package. In the model, we include the diffuse background components and all 4FGL-DR4 sources within the ROI, except for 4FGL J0851.9-4620e. 
The TS maps of RX J0852.0-4622 in the 5--500\,GeV and 20--500\,GeV energy ranges are shown in the left and right panels of Fig.~\ref{fig:cmap}, respectively.
As illustrated in the TS maps (Fig.~\ref{fig:cmap}), we also compare the \gray\ emission observed with H.E.S.S. (shown as contours) and with \fermi. The apparent difference between the two GeV maps is likely dominated by the broader LAT PSF at lower energies and by the Gaussian smoothing applied to the TS maps.

The TS maps show significantly enhanced GeV \gray\ emission from the location of the remnant, which aligns well with the shape of the remnant in the TeV energy range \citep{2018HESS}. 
The only significant discrepancy is likely attributable to the presence of the associated pulsar and its PWN within the extent of the remnant.
Whereas the Fermi-LAT data show a single bright feature in the southeast, the H.E.S.S. observations reveal two distinct regions of enhanced $\gamma$-ray emission in this area. 
The additional southeastern component, which lacks a counterpart in the LAT map, is plausibly associated with the pulsar/PWN system.

To characterise the GeV morphology, we perform binned likelihood fits to the 5--500\,GeV data using different spatial templates, and we compare these models using the Test Statistic (TS) and the Akaike Information Criterion (AIC, \citet{Akaike1974}). The corresponding best-fit parameters and fit statistics are summarised in Table~\ref{tab:likelihood}. The TS is defined as $\rm TS = 2 (\log{\cal L}_{1} - \log{\cal L}_{0})$, where ${\cal L}_{1}$ and ${\cal L}_{0}$ are the maximum likelihood values for the background with the target source and without the target source (null hypothesis). The AIC is defined as AIC = $-2\log({\cal L}) + 2k$, where $k$ is the number of free parameters in the model and $\cal L$ is the likelihood value of the corresponding model. To compare the goodness of the fit in the different models, we calculate the $\rm \Delta AIC$, defined as the difference between the AIC value of model 1 and those of models 2-6. A larger $\rm \Delta AIC$ therefore indicates that the corresponding model provides a better fit than model 1. 

\subsubsection{The uniform disk template}
In order to obtain the best spatial template of the GeV \gray\ emission in this region, we first test the uniform disk model adopted by \citet{2017Ackermann}, which we refer to as model 1.
Here, we refer to the best-fit disk model obtained from this work as model 2. The centre position (R.A., Dec) and radius are allowed to vary freely during the fitting, and the resulting best-fit parameters are summarised in Table~\ref{tab:likelihood}.
%The best-fit centre is found at $\rm R.A.=132.94\deg \pm 0.02\deg$, $\rm Dec=-46.34\deg \pm 0.02\deg$, and the optimal disk radius is $R_{\rm disk} = 1.02^{\circ} \pm 0.01 ^{\circ}$.

\subsubsection{The Gaussian template}
We also test a two-dimensional Gaussian template (model 3). For model 3, we allow both the centroid and width to vary freely, and the best-fit parameters are summarised in Table~\ref{tab:likelihood}.
However, model 3 shows no improvement compared to model 2, as indicated by the worse TS and $\Delta$AIC values provided in Table~\ref{tab:likelihood}.

\subsubsection{The eROSITA template}
To evaluate the spatial correlation between X-ray and $\gamma$-ray emission, we use the public eROSITA (eRASS1) data \citep{2024Merloni} in the 1--8\,keV energy range. Compared with earlier pointed \xray\ observations of Vela Jr, the eROSITA eRASS1 data provide uniform wide-field coverage of the full remnant, and are therefore well suited for constructing an X-ray template of the whole shell for comparison with the GeV morphology. The skytile datasets 134135, 132138, 130135 and 137138 covering the Vela Jr SNR are selected. Following the recommended processes in the eSASS\footnote{\url{https://erosita.mpe.mpg.de/dr1/eSASS4DR1/}} (eROSITA Science Analysis Software System) cookbook, we use the evtool task to generate the 1--8\,keV X-ray counts map of the SNR. We then compute an effective exposure map with the expmap task and estimate the instrumental background contribution with the erbackmap task. The final eROSITA image is obtained after background subtraction and exposure correction.
%\footnote{\url{https://erosita.mpe.mpg.de/dr1/eSASS4DR1/eSASS4DR1_cookbook/}}
This map serves as the eROSITA spatial template (model 4) and is shown in Fig.~\ref{fig:cmap} as black contours.
%The S/N is given by $\rm N_{src}/\sqrt{(N_{src}+N_{bkg})}$, where $\rm N_{src}$ is the number of (background-subtracted) source counts and $\rm N_{bkg}$ is the total number of background counts (sky background + source).
%Since X-ray point sources are typically much brighter than the extended emission and may not be associated with the \grays, we remove them from the eROSITA template and perform the morphology analysis for model 4, thereby improving the overall fit quality. 
Model 4 performs well and is favored over models 1-3, suggesting a strong spatial correlation between the X-ray and \gray\ emission. 

\subsubsection{The H.E.S.S. template}
Earlier measurements showed that the \gray\ emission originates from a thin shell rather than a sphere, based on its radial profile \citep{2007Aharonian, 2018HESS}. 
To quantify the spatial correlation between the extended GeV \gray\ emission and the TeV shell-like structure detected by H.E.S.S., we use the H.E.S.S. significance map as a spatial template (model 5). 
The TeV spatial template, as constructed from the H.E.S.S. observations of the remnant, is taken from \citet{2018HESS}. 
Model 6 is based on the H.E.S.S. template with the PWN region ($\rm R.A.=133.85\deg, Dec=-46.65\deg$) masked using a radius of 0.3\deg. 
The fit quality improves compared to model 5, indicating a modest improvement once the TeV PWN contribution is excluded.
The calculated values of $\rm \Delta AIC$ are listed in Table~\ref{tab:likelihood}.
The template with the largest $\rm \Delta AIC$ is model 6, which is therefore adopted as the best-fit spatial model for the spectral analysis.

\begin{figure}
%\centering
\includegraphics[scale=0.35]{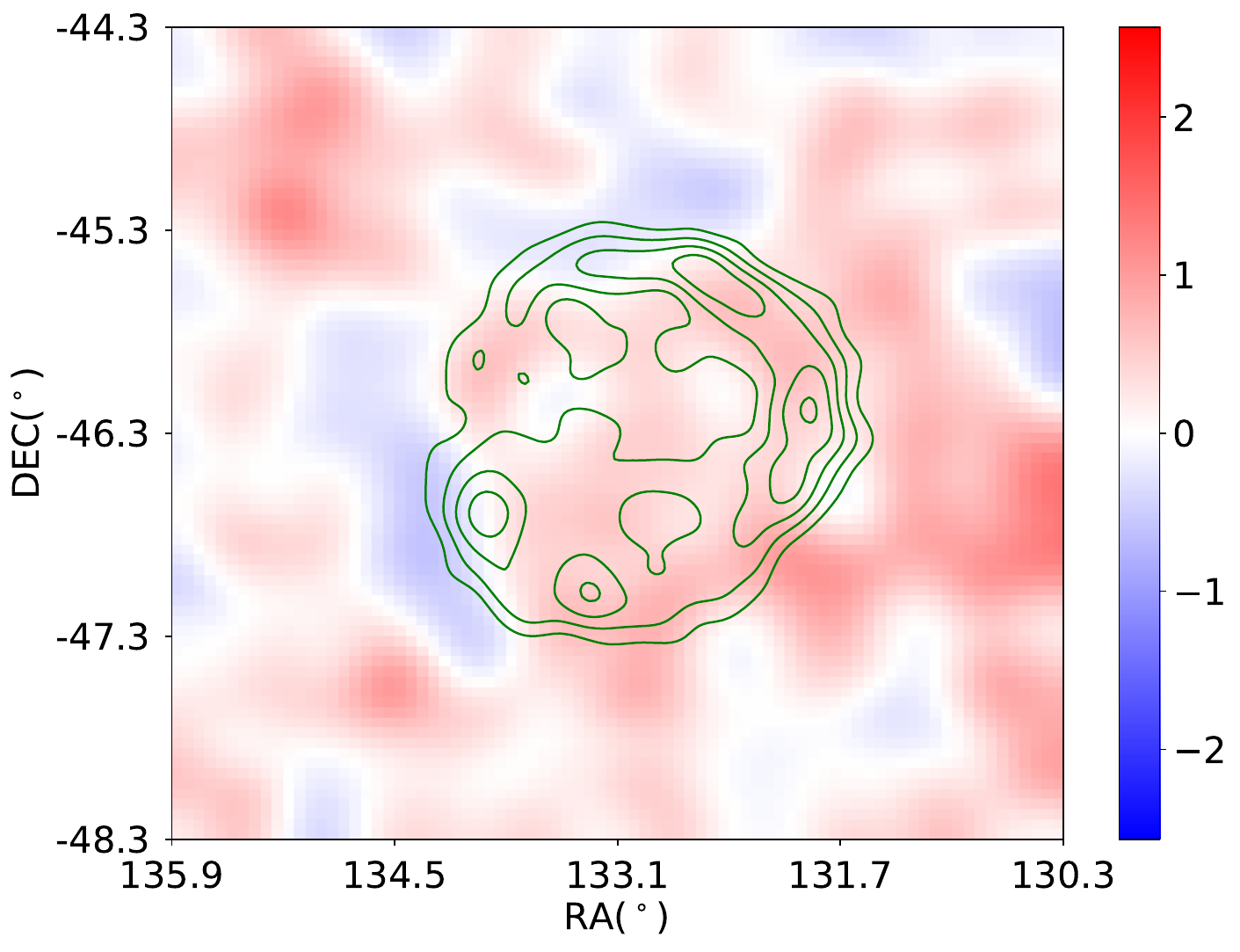}
\caption{The PS map for diagnostics of the goodness-of-fit, generated from \fermi\ data in the 5--500\,GeV range using the best-fit spatial model (model 6 in Table~\ref{tab:likelihood}). The colorbar limits are set between -2.57 and 2.57, corresponding to $\pm 3\sigma$. The green contours represent the same H.E.S.S. significance levels as those shown in white in Fig.~\ref{fig:cmap}}
\label{fig:allresdiual}
\end{figure}

To visually assess the goodness of fit of the adopted best-fit spatial model, we generate the Pseudo-Significance (PS) map\footnote{\url{https://fermi.gsfc.nasa.gov/ssc/data/analysis/user/gtpsmap/gtpsmap.py}} \citep{2021Bruel} using model 6, as shown in Fig.~\ref{fig:allresdiual}. The PS map reveals that the residuals are within $\pm 3\sigma$ across the region, indicating that the updated spatial model does not excessively subtract background emission and that there is no apparent excess in the remnant's surroundings.

%---------------------------------------------------- Figure 1plus for residual

\subsubsection{Azimuthal profile analysis}
%----------------------------------------------------- FIGURE 3
\begin{figure*}
%\centering
\includegraphics[scale=0.40]{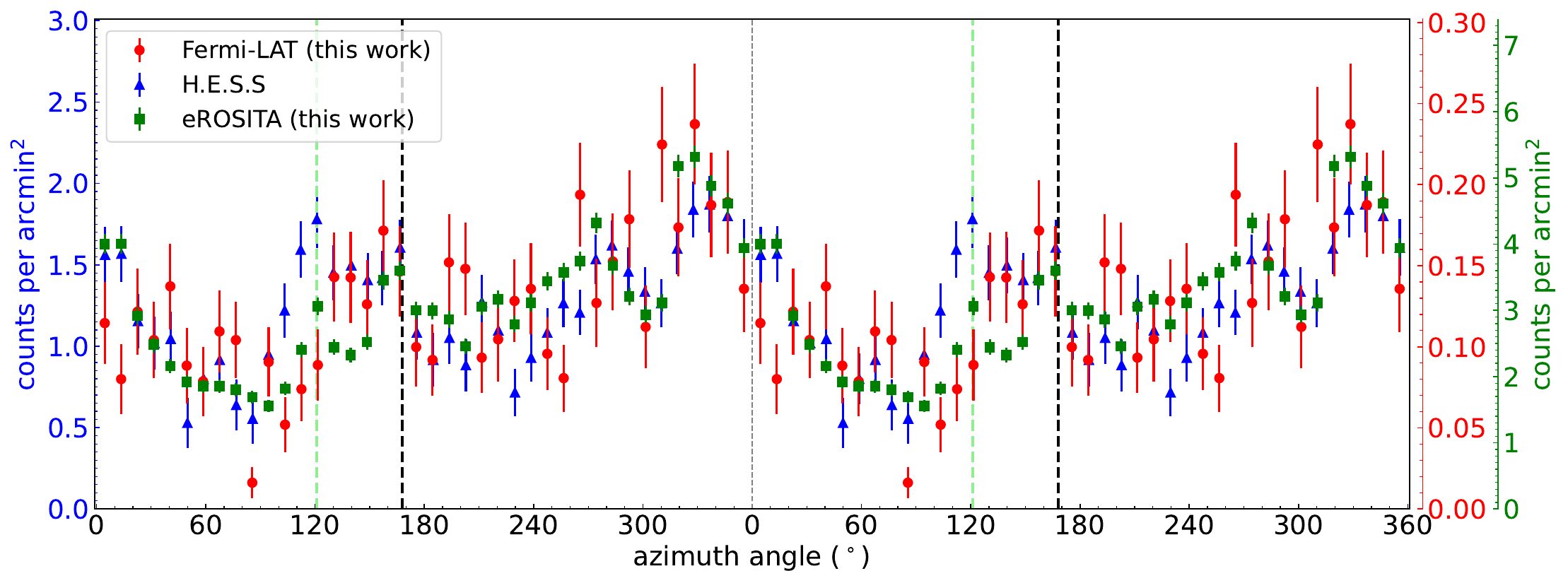}
\caption {The azimuthal profile extracted from the annulus in the skymap, shown as red points for the \fermi\ data (right scale), blue for the H.E.S.S. data (left scale), and green for the eROSITA data (right outer scale).
To facilitate comparison, a grey dashed line has been drawn between the two azimuthal periods. 
%The profiles are plotted over two periods, and a grey dashed line has been drawn between the two azimuthal periods to facilitate comparison.
The azimuthal position of the PSR J0855-4644 and the centre of the region around the southern enhancement detected by H.E.S.S. are indicated by green and black vertical dashed lines, respectively. 
}
\label{fig:profile}
\end{figure*}

We use the unsmoothed \fermi\ residual map in the 5--500\,GeV energy range to calculate photon counts per unit solid angle. 
To achieve a direct comparison with the H.E.S.S. data \citep{2018HESS}, the azimuthal profile is calculated from the photons in an annulus with inner and outer radii of $0.6 \deg$ and $1 \deg$, respectively, around the centre of RX J0852.0-4622, as shown by the red points in Fig.~\ref{fig:profile}. 
%The azimuthal profile for the same region in the H.E.S.S. observations, for energies above 100 GeV, is derived as well (blue points in Fig.~\ref{fig:profile}).
For the H.E.S.S. data, we adopt the azimuthal profile computed by \citet{2018HESS} for the same annular region and for energies above 100 GeV; the corresponding data points from their published profile are shown as the blue points in Fig.~\ref{fig:profile}.
We extract the azimuthal profile from the eROSITA counts map in 1--8\,keV energy range within the same annular region, which is shown as green points in Fig.~\ref{fig:profile}. 
The azimuth angle is deﬁned counterclockwise from the north. Two periods are separated by a dashed grey line. 
The green and black dashed lines at $121 \deg$ and $168 \deg$, respectively, denote the position of PSR J0855-4644 and the centre of the TeV emission enhancement seen toward the south of the shell. 

These azimuth profiles clearly show that both the GeV and TeV \grays\ emission are inhomogeneous along the shell, as expected from the corresponding \gray\ maps. The eROSITA X-ray profiles also exhibit a similar azimuthal modulation.
We find that the northwestern part of the shell (from $220\deg$ to $360\deg$) shows higher counts than the southeastern part in all three datasets.
Notably, a significant \gray\ emission enhancement is detected by H.E.S.S. between the southern ($\sim 160\deg$) and southeastern ($\sim 120\deg$) sectors \citep{2018HESS}, which is also observed in the GeV data, while the eROSITA profiles show a comparable structure. 
However, a less pronounced enhancement towards PSR J0855-4644 ($\sim 121\deg$) is observed in the case of GeV \grays, compared to the \xray\ and TeV \grays. 
The overall profiles of the three instruments show broadly consistent trends, confirming that the shell-like morphology dominates the emission from the X-rays to \gray, while local variations reflect energy-dependent structures along the rim. To quantify, we compute Pearson correlation coefficients over 0\deg--360\deg using the counts per unit solid angle in each azimuth bin, and obtain $r_{\rm LAT-HESS}=0.48$, $r_{\rm LAT-eROSITA}=0.60$, and $r_{\rm HESS-eROSITA}=0.73$, indicating that all three profiles are positively correlated and that the H.E.S.S.–eROSITA count profiles follow each other most closely.

\subsection{Spectral analyses}
\label{sec:spectral_analy}
To refine the spectral shape of the remnant's GeV counterpart by using model 6, we perform likelihood fits for various spectral models in the 0.1--500\,GeV energy range. The models tested include PowerLaw (PL), LogParabola (LogP), PLSuperExpCutoff (PLEC), and BrokenPowerLaw (BPL). The formulae and free parameters of these spectral models are presented in Table~\ref{tab:form}.
Table~\ref{tab:form} shows the fitting results, including the number of free parameters in the  likelihood model within ROI and $\Delta \rm AIC$ for each model.
%The BPL model, which has two more parameters than the simple PL model, fits the data slightly better, but the improvement is minimal. The LogP and PLEC models fit worse, even though they have more parameters. 
Thus, the PL best describes the $\gamma$-ray spectral shape. 
We derive that the photon spectral index of the masked H.E.S.S. template with a single PL spectrum is $1.77 \pm 0.03$, which is consistent within uncertainties with earlier estimates by \citet{2011Tanaka} and \citet{2017Ackermann}, and the total \gray\ flux in the 0.1--500\,GeV energy range is estimated as $(6.49 \pm 0.01_{\rm stat}) \times 10^{-8}\ \rm ph\ cm^{-2}\ s^{-1}$. 
%We adopt the distance $\sim 750$ pc for RX~J0852.0-4622, as proposed by \citet{2008Katsuda}, since the estimation only relies on the assumption of the shock speed, which has both theoretical and observational foundations. 
We adopt the distance $1.41\pm 0.14\ \rm kpc$  for RX~J0852.0-4622, as proposed by \citet{2025Suherli}.
Considering the distance, the total \gray\ luminosity is estimated to be $(2.47 \pm 0.49) \times 10^{33}\ \rm erg\ s^{-1}$.

\begin{figure}
%\centering
\includegraphics[scale=0.44]{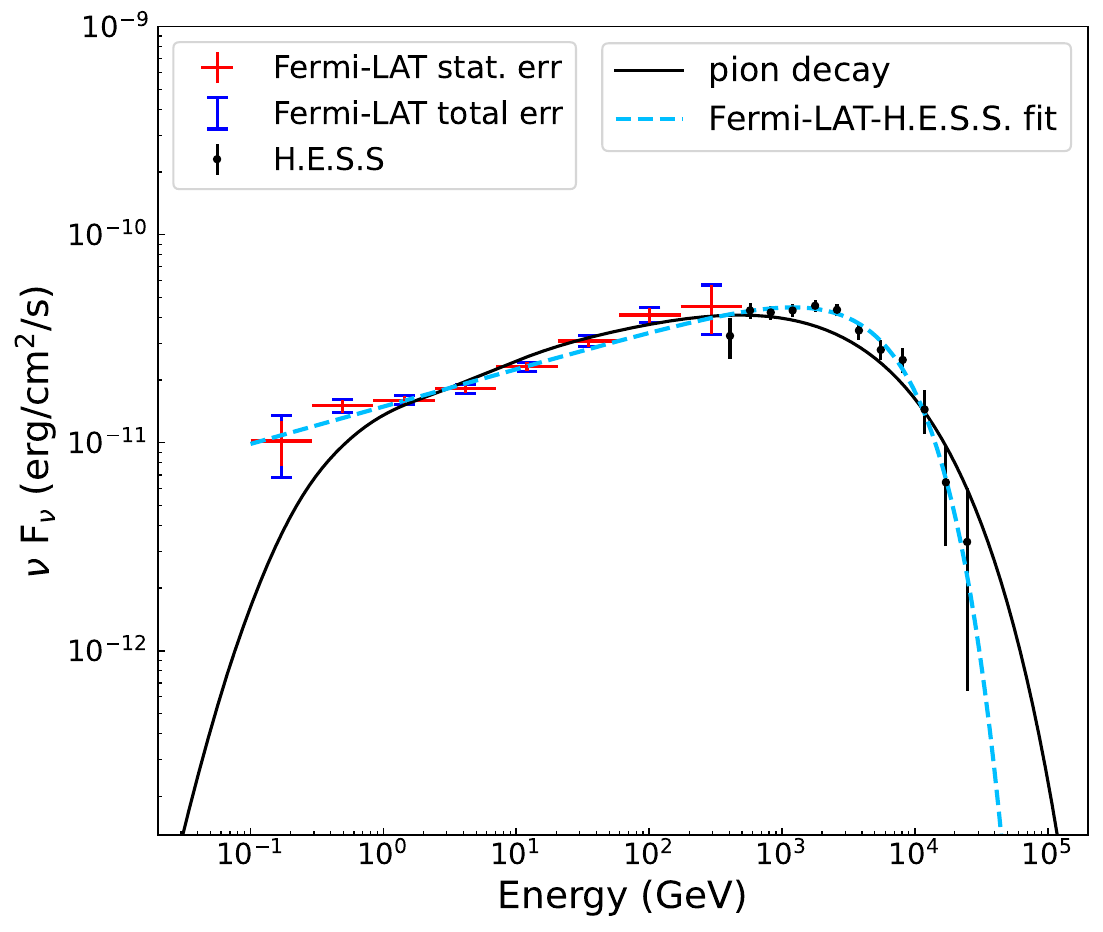}
\caption {SED of \gray\ emission towards RX J0852.0-4622 (red points) extracted from the H.E.S.S. template in the energy range from 100 MeV to 500 GeV by \fermi. Red error bars show statistical errors, and blue error bars represent the quadrature sum of statistical and systematic errors. The black data points represent the H.E.S.S. energy flux spectra taken from RX J0852.0-4622. The blue dashed line represents the simultaneous \fermi-H.E.S.S. fit using the ECPL.
%The expected sensitivity of the Cherenkov Telescope Array (CTA) \citep{2023HofmannCTA}, shown as a grey dash-dotted line. 
The solid curve represents the spectrum of \grays\ from interactions of relativistic protons with the ambient gas, assuming an exponential cutoff power-law distribution of protons (see sect.~\ref{sec:hadronic}).}
\label{fig:sed}
\end{figure}

We then perform the spectral analysis using the same best-fit template and adopting a PL spectral shape to extract the SED.
We divide the energy range of 0.1--500\,GeV into eight logarithmically spaced energy bins, and the SED flux in each bin is derived via the maximum-likelihood method. We calculate 95\% statistical errors for the energy flux densities.
The derived SED is shown in Fig.~\ref{fig:sed} as red points.
In the analysis, we estimate the systematic uncertainties of the SEDs due to the Galactic diffuse emission model and the LAT effective area by varying the normalization by $\pm 6\%$ from the best-fit value for each energy bin. We consider the maximum flux deviations of the source as the systematic error \citep{Abdo09a}. 
To account for the effect of spatial model selection, we extract the SED using the eROSITA template (model 4) and compare it to the result obtained from the best-fit template (model 6). The differences in each energy bin are included in the total systematic uncertainty.

We further perform a simultaneous fit to the combined \fermi\ and H.E.S.S. spectrum, adopting PL, LogP and PLEC models, where for the PLEC model we fix b=1. 
The PLEC model provides the best fit ($\chi^2/\rm ndf=16.46/17$) compared to the other two. 
The simultaneous \fermi-H.E.S.S. fit using the power law with exponential cutoff model is shown in Fig.~\ref{fig:sed} with a blue dashed line, and the fit parameters are shown in the right column of Table~\ref{tab:joint_fit}. 
Systematic uncertainties of the \fermi-H.E.S.S. fit are estimated following \citet{2018HESS}. \fermi\ points are shifted down (up), while H.E.S.S. points are shifted up (down), to test the impact on the spectral index and cutoff energy, and all points are shifted simultaneously in the same direction to evaluate the normalization. 
A comparison with the simultaneous \fermi-H.E.S.S fit reported by \citet{2018HESS} (Table~\ref{tab:joint_fit}) shows good agreement for all fit parameters.  

%----------------------------------------------------- FIGURE 4

\begin{table*}
\centering
\caption{Formulae for $\gamma$-ray spectral analysis and the corresponding fit results (0.1--500\,GeV) for different spectral models, using the best-fit spatial template (model 6).}
\label{tab:form}
\begin{tabular}{llllllll} 
\hline 
Name & Formula  &Free parameters &  $k$ & $\rm \Delta AIC$ \\ %$-\log(\mathcal{L})$  & 
\hline
PL   & $\frac{\mathrm{d}N}{\mathrm{d}E}$ = $N_0 {\left(\frac{E}{E_0}\right)}^{-\Gamma}$   & $N_0$, $\Gamma$  &  82 & 0 \\ %-18368120.4 & 
LogP & $\frac{\mathrm{d}N}{\mathrm{d}E}$ = $N_0 \left(\frac{E}{E_\mathrm{b}}\right)^{-\Gamma - \beta \log\left(\frac{E}{E_\mathrm{b}}\right)} $  & $N_0$, $\Gamma$, $\beta$ &  83 & -2\\ %-18368120.4 &
PLEC & $\frac{\mathrm{d}N}{\mathrm{d}E}$ = $ N_0 \left(\frac{E}{E_0}\right)^{-\Gamma_1} \exp\left({-\left({\frac{E}{E_\mathrm{cut}}}\right)^{\rm b}}\right)$  &  $N_0$, $\Gamma$, $E_\mathrm{cut}$, {\rm b} &  84 & -11\\ %-18368116.8 &
BPL &   $\frac{\mathrm{d}N}{\mathrm{d}E}$ = $\begin{cases} N_0{\left(\frac{E}{E_\mathrm{b}}\right)}^{-\Gamma_1} &  \mbox{: } E<E_\mathrm{b} \\ N_0\left(\frac{E}{E_\mathrm{b}}\right)^{-\Gamma_2} & \mbox{: }E>E_\mathrm{b} \end{cases}$ & $N_0$, $\Gamma_1$, $\Gamma_2$, $E_{\rm b}$ &  84 & -5 \\ %-18368120.0 &
\hline
\end{tabular}
\end{table*}

%\begin{table*}
%\centering
%\caption{Results of spectral analyses (0.1--500\,GeV) for different spectral models, using the best-fit spatial template (model 6). The curvature test statistic is defined as $\rm TS_{curve} = 2(log\cal L_{\rm curved} - \rm log\cal L_{\rm PL})$, where $\cal L_{\rm curved}$ is the likelihood of the curved model.}
%The significance for different spectral types ($\sigma_{\rm model}$) favoured over a PL null hypothesis in the 100 MeV - 500 GeV energy range.} %0.5--500\,GeV 
%\label{tab:spectral} 
%\begin{tabular}{lcccc}
%\hline
%Spectral Model    &  $-\log(\mathcal{L})$ & $\rm TS_{curve}$ & $k$ & $\rm \Delta AIC$ \\ %  & AIC & $\rm \Delta AIC$ $\Delta \log(\mathcal{L})$  \\
%\hline
%PL  &  -18368120.4 & - & 82 & 0\\%& -36736072 & -\\
%LogP  & -18368120.4 & 0 & 83 & -2\\%&  & -36736052 & 20 \\
%log(Likelihood) PL 80263.192 & BPL 80263.219  & LogPL 80264.589 & PLEC 80263.97& 5--500\,GeV
%$\sigma_{\rm model}$ (5--500\,GeV) PL 0 & LogP 1.67 & PLEC 1.19 & BPL 0.23
%log(Likelihood) PL 18368118.34 & BPL 18368122.28  & LogPL 18368110.10   & PLEC 18368114.67   & 0.1--500\,GeV
%PLEC & -18368116.8 & -7.2 & 84 & -11 \\  % & -36736062 & 10 \\ 
%BPL  & -18368120.0 &-0.8 & 84 & -5\\% & -36736076 &-4\\
%\hline
%\end{tabular}\\
%\end{table*}

\begin{table*}
\centering
\caption{Fit parameters of \fermi-H.E.S.S. spectral fit obtained in this work (right column), compared with the results reported by the \citet{2018HESS} (centre column).}
\label{tab:joint_fit} 
\begin{tabular}{lcccc}
\hline\hline
Parameter    &  \citet{2018HESS} & This work \\ 
\hline
$\Phi_{0}$ [$10^{-12} \rm cm^{-2} s^{-1} TeV^{-1}$] & $31.6 \pm 1.4_{\rm stat} \pm 7.6_{\rm syst}$ & $32.3 \pm 1.2_{\rm stat} \pm 3.1_{\rm syst}$\\
$\Gamma$ & $1.79 \pm 0.02_{\rm stat} \pm 0.10_{\rm syst}$ & $1.82 \pm 0.01_{\rm stat}\pm 0.02_{\rm syst}$\\
$E_{\rm cut}$\ [$\rm TeV$] & $6.6 \pm 0.7_{\rm stat} \pm 1.3_{\rm syst}$ &  $6.7 \pm 0.6_{\rm stat} \pm 1.0_{\rm syst}$\\
$E_{\rm 0}$\ [$\rm TeV$] &1 &1\\
$E_{\rm min}-E_{\rm max}$\ [$\rm TeV$] &0.001-30 &0.0001-30\\
$F(>1\ \rm TeV)$ [$10^{-12} \rm cm^{-2} s^{-1}$] & $23.2 \pm 0.7_{\rm stat} \pm 5.6_{\rm syst}$ & $23.3 \pm 0.6_{\rm stat}\pm 2.2_{\rm syst}$\\
$F(0.3-30\ \rm TeV)$ [$10^{-12} \rm cm^{-2} s^{-1}$] & $81.7 \pm 2.6_{\rm stat} \pm 19.6_{\rm syst}$ & $84.7 \pm 2.2_{\rm stat} \pm 8.0_{\rm syst}$\\
\hline
\end{tabular}\\
\end{table*}

%%%%%%%%%%%%%%%%%%%%%%%%%%%%%%%%%%%%%%%%%%%%%%%%%%%%%%%%%%%%%%%%%%%%%%%%%%%%%%%%
\section{The origin of gamma-ray emission}
\label{sec:origin}
In this section, we model the MWL SED of RX J0852.0$-$4622 using two broadband scenarios: a pure leptonic model and a hybrid lepton-hadron model. 
%In this section, we present the physical models that account for the spectrum of the GeV-TeV \gray\ emission and discuss its possible origin. 
The MWL data used in this section consist of Parkes radio data points from \citet{2000Duncan}, \xray\ data from the first eROSITA all-sky survey, \fermi\ GeV \gray\ points from this work (red points in Fig.~\ref{fig:sed} and \ref{fig:sed_IC}), and H.E.S.S. TeV \gray\ points (black points in Fig.~\ref{fig:sed} and \ref{fig:sed_IC}) from \citet{2018HESS}.
For the eROSITA data, we define a circular region with a radius of $1.0\deg$ centered at R.A.=$133.08\deg$, Dec=$-46.34\deg$ as the source region, and an annular region with inner and outer radii of $1.0\deg$ to $1.2\deg$, respectively, as the background region, while masking bright point sources. 
The spectra of the source and background, along with the redistribution matrix file (RMF) and ancillary response file (ARF), are extracted using the srctool task. We perform a spectral fit in XSPEC in 1--5\,keV band using the RMF/ARF and an absorbed power-law model to convert the spectra from count space to flux space.
%In Fig.~\ref{fig:sed_IC}, the eROSITA differential flux of the entire remnant is shown as green points, with statistical uncertainties calculated at the 68\% confidence level.
In Fig.~\ref{fig:sed_IC}, the eROSITA SED of the entire remnant is shown as green points. The spectrum is rebinned into five logarithmic sub-bands, and the statistical uncertainties are calculated at the 68\% confidence level.

%We fit the MWL data points using hadronic and leptonic scenarios to derive the particle distributions required using the Naima \footnote{\url{https://naima.readthedocs.io/en/latest/index.html}} \citep{2015Zabalza} package.
We use Naima \footnote{\url{https://naima.readthedocs.io/en/latest/index.html}} \citep{2015Zabalza} package and assume an exponential cutoff power-law parent particle spectrum, 
\begin{equation}
    N(E) = A~E^{-\alpha}\exp(- \frac{E}{E_{\rm cutoff}}),
\label{equ:ECPL}
\end{equation}  
where $A$, $\alpha$, and $E_{\rm cutoff}$ are free parameters. The purely hadronic $\gamma$-ray model shown in Fig.~\ref{fig:sed} is retained only as a $\gamma$-ray-only benchmark, and is not intended as a physically complete broadband interpretation.

\subsection{Leptonic Scenario}
\label{sec:leptonic}
%----------------------------------------------------- TABLE 3
\begin{table*}
\centering
        \caption{Best parameter sets of the parent particle distribution, assuming an exponential cutoff power-law in the leptonic and hybrid scenarios.}
\begin{tabular}{ccccccccc}
\hline
        &Scenario & norm [1/eV]  & $\alpha$ & $E_{\rm cut}$ (TeV) &  B ($\mu$G) & $W (\rm erg)$    & $\Delta \rm AIC$ \\ 
\hline
        %&Hadronic & $(3.1 \pm 0.1) \times 10^{35}$ & $1.84 \pm 0.01$ & $67\pm 7$ &  $(4.8\pm 0.1)\times 10^{47}$  & - & -23.0 \\ %\cline{1-1}
        &Leptonic & $(1.83\pm0.04) \times 10^{35}$ & $2.23 \pm 0.02$ & $21.8\pm1.2$ & $6.88 \pm 0.19$ & $(5.5\pm0.3) \times 10^{48}$  &  0 \\ %\cline{1-1}
        &Hybrid & $\rm A_{e}=(1.26 \pm 0.06) \times 10^{35}$ & $2.18 \pm 0.02$ &  $E_{\rm cut,e}=23.4\pm 1.3$  & $6.9 \pm0.2$ & $W_{\rm e}=(3.2\pm 0.3)\times 10^{48}$  &  204 \\
        &  & $\rm A_{p}=(6.4 \pm 0.6) \times 10^{35}$& & $E_{\rm cut,p}= 60\pm 8$  & & $W_{\rm p}=(1.66\pm 0.11)\times 10^{49}$   \\
\hline
\end{tabular}
\label{table:origin}
\end{table*}

A pure leptonic model provides the simplest broadband description once the radio and X-ray synchrotron emission is taken into account. In this model, the radio to X-ray emission is produced by synchrotron radiation from relativistic electrons, while the GeV--TeV $\gamma$-ray emission is generated via inverse Compton (IC) scattering.
For the photon field of the IC calculations, the cosmic microwave background (CMB), the optical-UV radiation from starlight, and the dust infrared (IR) radiation field might also be considered. 
According to the interstellar radiation field model by \citet{2006Porter}, in observations of shell-type SNRs in the outer Galaxy, the CMB photons provide the majority of the IC emission. 
Thus, the contribution of IR and optical radiation fields should be negligible due to the large distance between the SNR RX J0852.0-4622 and the Galactic centre ($\sim9$ kpc).
In this case, we only consider the CMB as the radiation field responsible for the IC scattering of relativistic electrons. We calculate the IC spectrum using the formalism described in \citet{2014Khangulyan}.

We assume an exponential cutoff power-law distribution (same function as  Eq.~\ref{equ:ECPL}) of the relativistic electrons. 
For the leptonic models, the magnetic field and the exponential energy cutoff are treated as independent parameters. 
The magnetic field is constrained by the ratio of synchrotron to IC flux magnitude, whereas the cutoff in the parent electron spectrum is constrained by the cutoff in the very-high-energy part of the IC \gray\ spectrum.
%The best-fitting results for electrons above 1 GeV are $\alpha = 2.38 \pm 0.02$, $E_{\rm e,cut} = 23.9^{+2.0}_{-1.6}$ TeV, and $W_{\rm e} = (3.6^{+0.4}_{-0.3}) \times 10^{48}\ \rm erg$, with a maximum log-likelihood value of -31.5.
The best-fitting leptonic SED is shown in the top panel of Fig.~\ref{fig:sed_IC}, and the corresponding parameters are listed in Table~\ref{table:origin}. The corresponding synchrotron fluxes, computed assuming an average magnetic field of B = 6.8 $\mu$G, are also shown in Fig.~\ref{fig:sed_IC}.
The derived magnetic field aligns well with previous leptonic models \citep{2023Camilloni, 2018HESS, 2013Lee} and can be interpreted as an average magnetic field across the SNR shell. This does not rule out the possibility of regions with either higher or lower magnetic field strengths. According to \citet{2013Lee}, such a low magnetic field suggests the presence of a wind-blown cavity, which is necessary to maintain a weak magnetic field in the upstream medium.
We note that the derived energy budget for relativistic electrons ($>$ 1 GeV) is $\sim (5.5\pm0.3) \times 10^{48} \rm erg$, corresponding to an electron energy fraction $\eta_{e}= W_{\rm e} / E_{\rm SN} \sim 0.6\%$ for the typical kinetic energy of a supernova explosion ($\sim 10^{51}\ \rm erg$).
%We note that the derived energy budget for relativistic electrons ($>$ 1 GeV) is $\sim 5.5 \times 10^{48} \rm erg$, which is significantly lower than the typical kinetic energy of a supernova explosion ($\sim 10^{51}\ \rm erg$). This indicates that the system has sufficient power to account for the detected $\gamma$-ray emission. 

\subsection{Hybrid Scenario}
\label{sec:hadronic}
\begin{figure}
\includegraphics[scale=0.36]{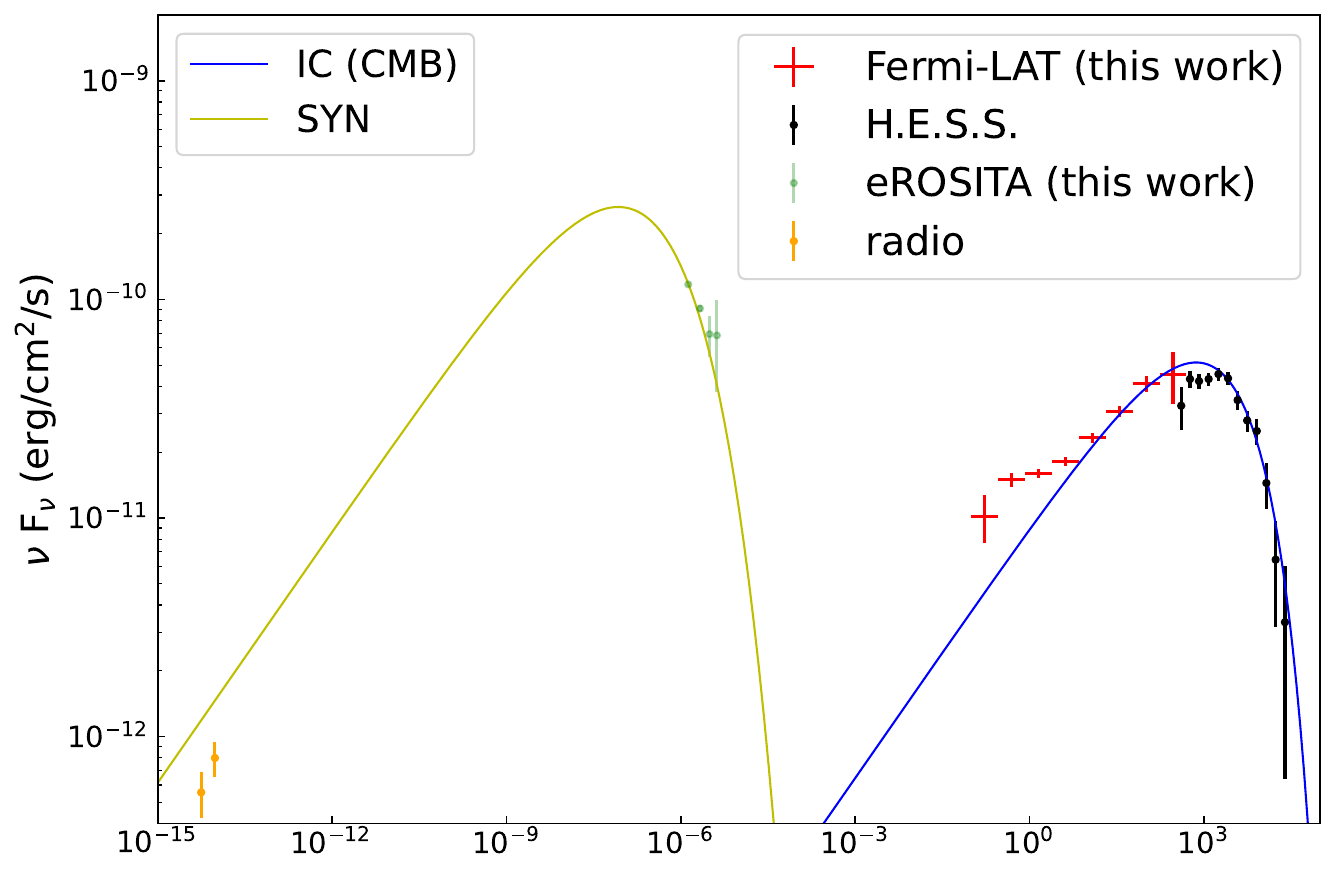}
\includegraphics[scale=0.36]{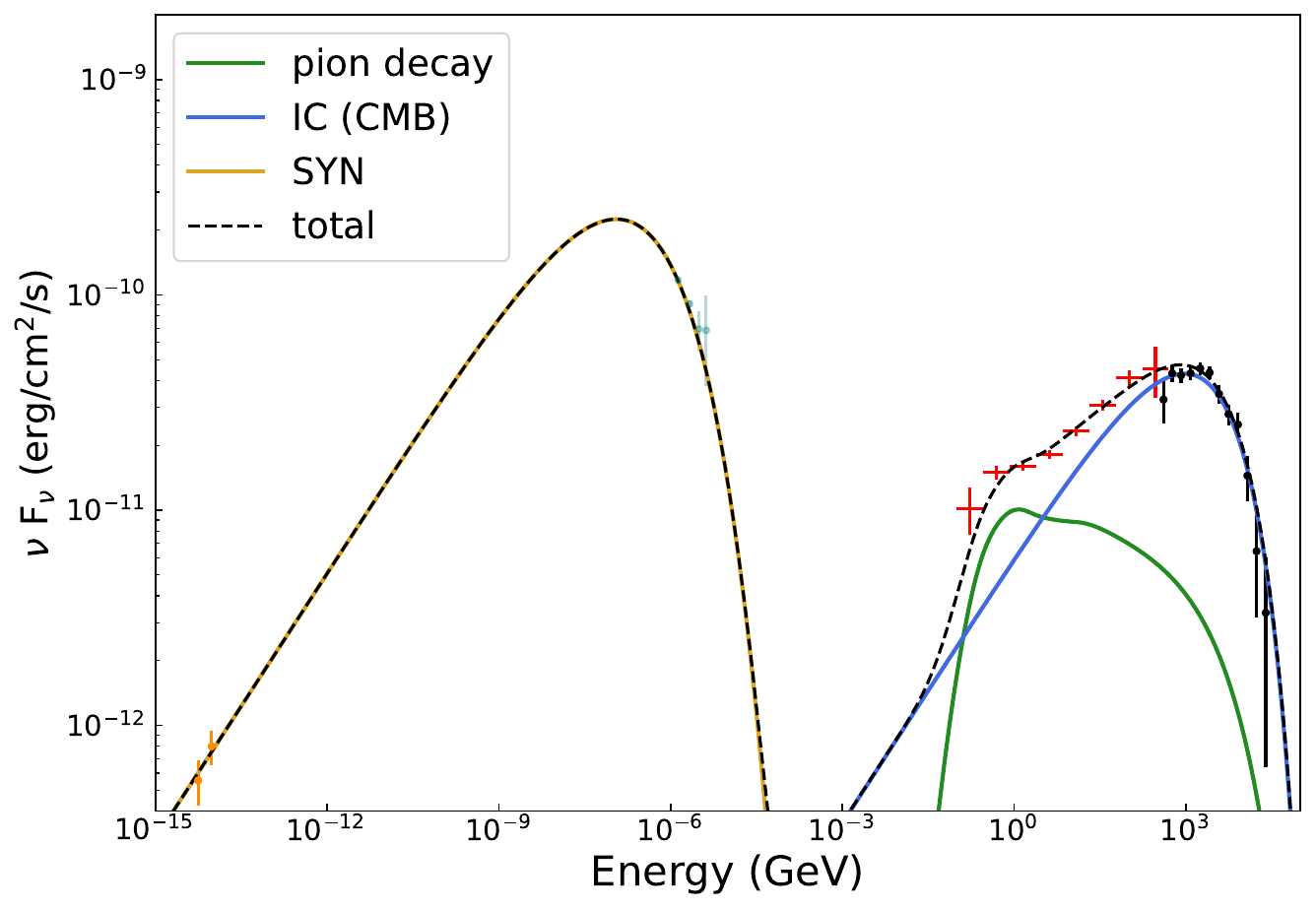}
\caption {
The broadband SED of RX J0852.0-4622 with the leptonic (top) and hybrid (bottom) scenarios (see sect.~\ref{sec:origin}). The radio data, shown as orange points, were adopted by \protect\citet{2000Duncan}. The green points represent the eROSITA flux extracted in the 1--5\,keV band from the entire remnant, with statistical uncertainties calculated at the 68\% confidence level. GeV--TeV \gray\ data are the same as Fig.~\ref{fig:sed}. The yellow solid line shows the synchrotron component for RX J0852.0-4622. The blue solid curve represents IC scattering of CMB seed photons, and the green solid curve denotes the $\pi^{0}$-decay emission from hadronic interactions.}

\label{fig:sed_IC}
\end{figure}

Although the radio and X-ray synchrotron emission clearly requires relativistic electrons, the spatial correlation between the $\gamma$-ray emission and the molecular and atomic gas around RX J0852.0$-$4622 \citep{2017Fukui}, together with the gas distribution shown in Appendix~\ref{sec:Gas}, suggests that a hadronic component may also be present. This mixed-origin picture is further motivated by the recent spatial decomposition study of \citet{2024Fukui}, who quantified the hadronic and leptonic $\gamma$-ray components in RX J0852.0$-$4622 and found an approximately equal hadronic-to-leptonic ratio ($\sim 5:5$) in $\gamma$-ray counts. Their work emphasizes that the target interstellar protons, in particular their spatial distribution, are essential for identifying the origin of the $\gamma$-ray emission.
We therefore test a hybrid lepton-hadron model, in which the radio and X-ray emission is attributed to synchrotron radiation from relativistic electrons, while the observed GeV--TeV $\gamma$-ray emission is modeled as the sum of leptonic IC and hadronic $\pi^0$-decay. In the hadronic component, we adopt the average target proton density of $n_{\rm H}=5.8~{\rm cm^{-3}}$, derived from the gas distributions in Appendix~\ref{sec:Gas} considering $\rm H_2$ + HI gas. Assuming charged particles share the same acceleration mechanism, the spectral index $\alpha$ of protons could be identical to that of electrons \citep{2004Petrosian, 2011Yuan}. 
The best-fitting hybrid SED is shown in the bottom panel of Fig.~\ref{fig:sed_IC}, and the corresponding parameters are listed in Table~\ref{table:origin}. 
A comparison with the pure leptonic model shows that the hybrid scenario provides a better statistical description of the same MWL dataset, with $\Delta$AIC = 204 in favour of the hybrid fit. As shown in Fig.~\ref{fig:sed_IC}, the hybrid model retains the synchrotron interpretation of the radio--X-ray emission and a predominantly leptonic IC origin of the TeV emission, while the additional hadronic $\pi^0$-decay component improves the description of the GeV band.
In this model, the TeV emission is still dominated by the leptonic IC component, while the hadronic $\pi^0$-decay component contributes mainly through the characteristic pion bump at a few hundred MeV, where the modeled hadronic emission is several times brighter than the leptonic IC component.

By integrating the model energy fluxes, we find that in the 0.1--300 GeV band the IC and $\pi^0$-decay components contribute 66\% and 34\%, respectively, while in the 0.3--30 TeV band the corresponding contributions are 92\% and 8\%. Therefore, our SED modeling supports a mixed-origin picture in which the GeV emission includes a non-negligible hadronic contribution, whereas the TeV band remains predominantly leptonic. Although the ratios derived here refer to band-integrated energy fluxes, rather than the $\gamma$-ray counts used by \citet{2024Fukui}, both approaches consistently indicate coexisting leptonic and hadronic components in RX J0852.0$-$4622.

\section{Discussion and Conclusions}
\label{sec:Disc}
Multi-band observations demonstrate the complex nature of this region, which includes the shell-type SNR Vela Jr, the bright pulsar PSR J0855-4644, and its associated PWN, as well as the nearby Vela SNR lying in the foreground, further complicating the region.
In this work, we use more than 15 yr of Fermi-LAT data, show that the GeV morphology is best described by the masked H.E.S.S. shell template, construct an independent eROSITA shell template and 1--5 keV SED for the whole remnant, and perform a direct broadband comparison between the pure leptonic and hybrid lepton-hadron models using the same MWL dataset. The existence of an extended GeV shell and its hard spectrum smoothly connecting to the TeV band confirm previous results, while the improved GeV spatial modeling and the new eROSITA constraints are the main additions of this work.
Previous studies of the TeV pulsar population, such as \citet{2018HESS_PWN} and \citet{2026HESSPWN}, suggest that PWNe of energetic pulsars are likely detectable in TeV \grays.
The azimuthal profile from H.E.S.S. observations (blue points in Fig.~\ref{fig:profile}) indeed shows that the flux at the position coinciding with the PWN is approximately double that of the surrounding region, whereas the \fermi\ data (red points in Fig.~\ref{fig:profile}) show no significant variation at the pulsar position.
According to the tests with PWN-masked H.E.S.S. templates, the fit quality improves. Based on these results, we argue that PSR J0855-4644 and its associated PWN likely do not significantly contribute to the extended GeV \gray\ source. 
The complexity of the region and the limited understanding of the interstellar medium introduce uncertainties when modeling the Galactic diffuse \gray\ background. The residuals may be due to imperfect modeling of the diffuse background, especially given the spatial overlap of \grays\ and surrounding gas. As illustrated in Fig.~\ref{fig:sed}, the hard spectrum of this region (with an index of 1.77) is inconsistent with the softer Galactic diffuse \gray\ background, which has an index of 2.7.

The \fermi\ data analysis presented in this work shows that the GeV counterpart of Vela Jr is spatially coincident with the TeV excess detected by H.E.S.S., namely HESS J0852-463. 
A detailed analysis of \fermi\ data shows that the hard GeV spectrum can smoothly connect to the spectrum in the TeV band detected by \citet{2018HESS}. 
%The spatially resolved spectroscopy study by \citet{2018HESS} shows no significant TeV spectral variation across the SNR, implying that the parent particle population is the same throughout the remnant and that the conditions for particle acceleration are similar everywhere.    
The spatially resolved spectroscopy study by \citet{2018HESS} shows no significant TeV spectral variation across the SNR. 
In our work, the global GeV morphology is broadly consistent with both the TeV shell and the eROSITA shell, but the current LAT angular resolution and photon statistics do not allow a spatially resolved GeV spectroscopy comparable to the H.E.S.S. analysis. 

%Therefore, the physically relevant broadband comparison is not a purely hadronic versus purely leptonic dichotomy, but rather the relative importance of leptonic and hadronic components. Since SNRs can also be natural \gray\ emitters, both hadronic and leptonic scenarios, as mentioned above, can account for the \gray\ emission. \citet{2021Zeng} argued that the GeV \gray\ emission originates from part of the SNR shell, and suggested that the old age of the SNR would rule out a leptonic origin for the TeV \gray\ emission. In this work, we cannot formally rule out a leptonic origin for this source only by considering the age of Vela Jr.
\citet{2021Zeng} argued that the GeV \gray\ emission originates from part of the SNR shell, and suggested that RX J0852.0$-$4622 can be described by both leptonic and hadronic models with reasonable parameters.
The presence of synchrotron emission in the radio and X-ray bands already demonstrates that relativistic electrons are present in the shell. In addition, the high forward-shock velocity of Vela Jr \citep{2025Suherli} indicates that the remnant is still capable of accelerating electrons to multi-TeV energies.

\citet{2017Fukui} found a good spatial correspondence between the TeV \grays\ from Vela Jr and the interstellar protons from molecular and atomic line observations, further supporting a hadronic component in the \grays\ from this SNR.
Our broadband modeling in sect.~\ref{sec:origin} shows that the hybrid lepton-hadron scenario provides a natural mixed-origin interpretation, in which the hadronic contribution is mainly relevant in the GeV band, while the TeV emission remains predominantly leptonic.
By integrating the model energy fluxes, we obtain IC/$\pi^0$-decay contributions of 66\%/34\% in the 0.1--300 GeV band, and 92\%/8\% in the 0.3--30 TeV band. \citet{2024Fukui} decomposed the H.E.S.S. \gray\ counts above 100 GeV spatially into hadronic and leptonic components, using the interstellar proton column density and the Suzaku non-thermal X-ray counts as templates, and obtained an approximately equal hadronic/leptonic contribution for RX J0852.0-4622. Their estimate is based on spatially decomposed counts, whereas our result is based on band-integrated model energy fluxes derived from a broadband SED fit. The proton energy budget exceeds that of electrons by a factor of about 5, which is consistent with the general expectation from studies of individual SNRs and from the Galactic cosmic-ray spectrum that SNRs accelerate protons more efficiently than electrons \citep{2013Blasi}.

The derived total CR proton energy of about $10^{49}$ erg is also consistent with SNR scenarios. 
Since a shell is clearly resolved in $\gamma$-rays, it is more appropriate to use the shell thickness, rather than the full shell radius, as the characteristic diffusion length. The H.E.S.S. morphology of RX J0852.0$-$4622 indicates a shell extending from $0.6^\circ$ to $1.0^\circ$, implying a thickness of about 40\% of the outer radius \citep{2018HESS}. For a shell radius of $\sim 25$ pc at 1.41 kpc, this gives $l \sim 10$ pc. Using $D \sim l^2/4T$ with $T \lesssim 4300$ yr, we obtain $D \sim 1.8 \times 10^{27}\ {\rm cm^2\ s^{-1}}$. This is still below the canonical Galactic-plane diffusion coefficient, $D_{\rm gal}(E) \approx 3 \times 10^{28} (E/10\,{\rm GeV})^{\delta}\ {\rm cm^2\ s^{-1}}$ \citep{2013Blasi}, suggesting suppressed diffusion and efficient confinement of CRs within the shell.
However, morphology alone does not yet uniquely distinguish between the leptonic and hybrid interpretations.
Future multi-wavelength observations will be necessary to better constrain its nature.

In conclusion, we confirm the detection of GeV \gray\ emission toward the Vela Jr region, which is a shell-type SNR in our Galaxy. 
We find that the extended GeV \gray\ emission is best modeled by the H.E.S.S. spatial template rather than by regular geometrical shapes.
The GeV \gray\ emission reveals a hard spectrum that can be described by a power-law function with a photon index of about $1.77 \pm 0.03$. 
The GeV \gray\ characteristics of Vela Jr are similar to those of several shell-type SNRs, such as RX J1713.7-3946 \citep{2018HESS_RXJ1713} and HESS J1731-347 \citep{2018Guo, 2017Condon}. 
%The refined GeV \gray\ spectrum can smoothly connect to the H.E.S.S. measurements \citep{2018HESS} at TeV energies.
We confirm this result and further reduce systematic uncertainties in the joint spectral fit.
The new eROSITA analyses provide an independent shell-like \xray\ template consistent with the GeV morphology and tightly constrain the synchrotron emission of the remnant.
%The $\gamma$-ray spectrum of RX J0852.0-4622 can be described by both hadronic and leptonic scenarios. The leptonic scenario reproduces the MWL data with synchrotron and IC emission from electrons. The hadronic model, fitting only to \grays, attributes the emission to $\pi^{0}$ decay but lacks MWL consistency. Within the current statistical and systematic uncertainties, our modeling does not allow us to statistically distinguish between these two limiting scenarios, and both remain viable interpretations.
For the broadband MWL interpretation, the relevant comparison is between the pure leptonic and hybrid lepton-hadron scenarios. The pure leptonic model reproduces the MWL data with synchrotron and IC emission from electrons, while the hybrid model allows an additional hadronic contribution mainly in the GeV band. 
Within the current statistical and modeling uncertainties, the broadband data strongly favor a non-negligible hadronic contribution in the GeV band, although the exact hadronic fraction remains uncertain.

\section{Acknowledgements}
We thank Yun-Feng Liang and Xiao-Na Sun for useful discussions. 
This work is supported in part by the Guangdong Provincial Key Laboratory of Advanced Particle Detection Technology (2024B1212010005), the Guangdong Provincial Key Laboratory of Gamma-Gamma Collider and Its Comprehensive Applications (2024KSYS001), the Fundamental Research Funds for the Central Universities, and the Sun Yat-sen University Science Foundation.
This work is supported by the National Natural Science Foundation of China (NSFC) grant 12273122, 12205388, National Astronomical Data Center, the Greater Bay Area, under grant No. 2024B1212080003, and science research grant from the China Manned Space Project under CMS-CSST-2025-A13.
% a science research grant from the China Manned Space Project (No. CMS-CSST-2021-B11). 
This work is supported by a scholarship from the China Scholarship Council (CSC).

This work is based on data from eROSITA, the soft \xray\ instrument aboard SRG, a joint Russian-German science mission supported by the Russian Space Agency (Roskosmos), in the interests of the Russian Academy of Sciences represented by its Space Research Institute (IKI), and the Deutsches Zentrum für Luft- und Raumfahrt (DLR). The SRG spacecraft was built by Lavochkin Association (NPOL) and its subcontractors, and is operated by NPOL with support from the Max Planck Institute for Extraterrestrial Physics (MPE). The development and construction of the eROSITA \xray\ instrument was led by MPE, with contributions from the Dr. Karl Remeis Observatory Bamberg \& ECAP (FAU Erlangen-Nuernberg), the University of Hamburg Observatory, the Leibniz Institute for Astrophysics Potsdam (AIP), and the Institute for Astronomy and Astrophysics of the University of Tübingen, with the support of DLR and the Max Planck Society. The Argelander Institute for Astronomy of the University of Bonn and the Ludwig Maximilians Universität Munich also participated in the science preparation for eROSITA. The eROSITA data shown here were processed using the eSASS software system developed by the German eROSITA consortium.

\section{Data availability}
The \fermi\ data used in this work are publicly available, and are provided online at the NASA-GSFC Fermi Science Support Center\footnote{\url{ https://fermi.gsfc.nasa.gov/ssc/data/access/lat/}}. 
The eROSITA data (eRASS 1) used in this work are publicly available\footnote{\url{https://erosita.mpe.mpg.de/dr1/erodat/skyview/skytile_search/}}.
The TeV data products used in this work are taken from the published H.E.S.S. results of \citet{2018HESS}, including the shell significance map used as the spatial template, the published azimuthal profile, and the TeV SED points.
For the molecular gas analysis, we use the CfA 1.2 m CO survey data \citep{2001Dame}, accessed via the LAMBDA archive,\footnote{\url{https://lambda.gsfc.nasa.gov/product/foreground/fg_wco_info.html}} to derive the H$_2$ column density.
%We made use of the CO data\footnote{\url{ https://lambda.gsfc.nasa.gov/product/foreground/fg_wco_info.html}} to derive the H$_{2}$ column density. 
The HI data are taken from the HI4PI\footnote{\url{http://cdsarc.u-strasbg.fr/viz-bin/qcat?J/A+A/594/A116}}.

%%%%%%%%%%%%%%%%%%%%%%%%%%%%%%%%%%%%%%%%%%%%%%%%%%
\bibliographystyle{mnras}
%\bibstyle{aa}
\bibliography{ms}

\appendix
%%%%%%%%%%%%%%%%%%%%%%%%%%%%%%%%%%%%%%%%%%%%%%%%%%%%%%%%%%%%%%%%%%%%%%%%
%\section{Gas Observation}
\section{Gas Observation}
\label{sec:Gas}
The spatial correlation between \grays\ and interstellar protons is key to determining the \gray\ production mechanism \citep{2012Inoue,2012Fukui}.
According to the study by \citet{2017Fukui}, the good spatial correspondence between the \gray\ and the interstellar gas supports a hadronic component to the observed \gray\ emission. They also include the very large rim (which is observed in the 1--20 km/s velocity range as stated by \citet{2017Fukui}, which is not necessarily associated with the SNR and it clearly dominates the gas emission in the surrounding regions.
Consequently, the spatial distribution of \grays\ would generally follow that of the gas. Therefore, we investigate the spatial distribution of molecular hydrogen (H$_{2}$) and neutral atomic hydrogen (HI) around Vela Jr.

%----------------------------------------------------- TABLE 2
\begin{table}
\centering
        \caption{Total gas masses and number densities within the disc with a radius of $1.02\deg$.
        % See Appendix \ref{sec:Gas} for details.
        }
\begin{tabular}{cccc}
\hline
        Tracer & Mass ($\rm{10^{4}\msun}$) & Number density ($\rm {cm^{-3}}$) \\
\hline
        %$\rm H_{2}$ & 0.9 & 37 \\ %\cline{1-1}
        %HI & 2.9 & 118 \\ %\cline{1-1}
        %Total & 3.8 & 155\\ %\cline{1-1}\
        $\rm H_{2}$ & 3.2 & 1.4 \\ %\cline{1-1}
        HI & 10.1 & 4.4 \\ %\cline{1-1}
        Total & 13.3 & 5.8\\ %\cline{1-1}
\hline
\end{tabular}
\label{table:gas}
\end{table}

\begin{figure*}
\centering
\includegraphics[scale=0.3]{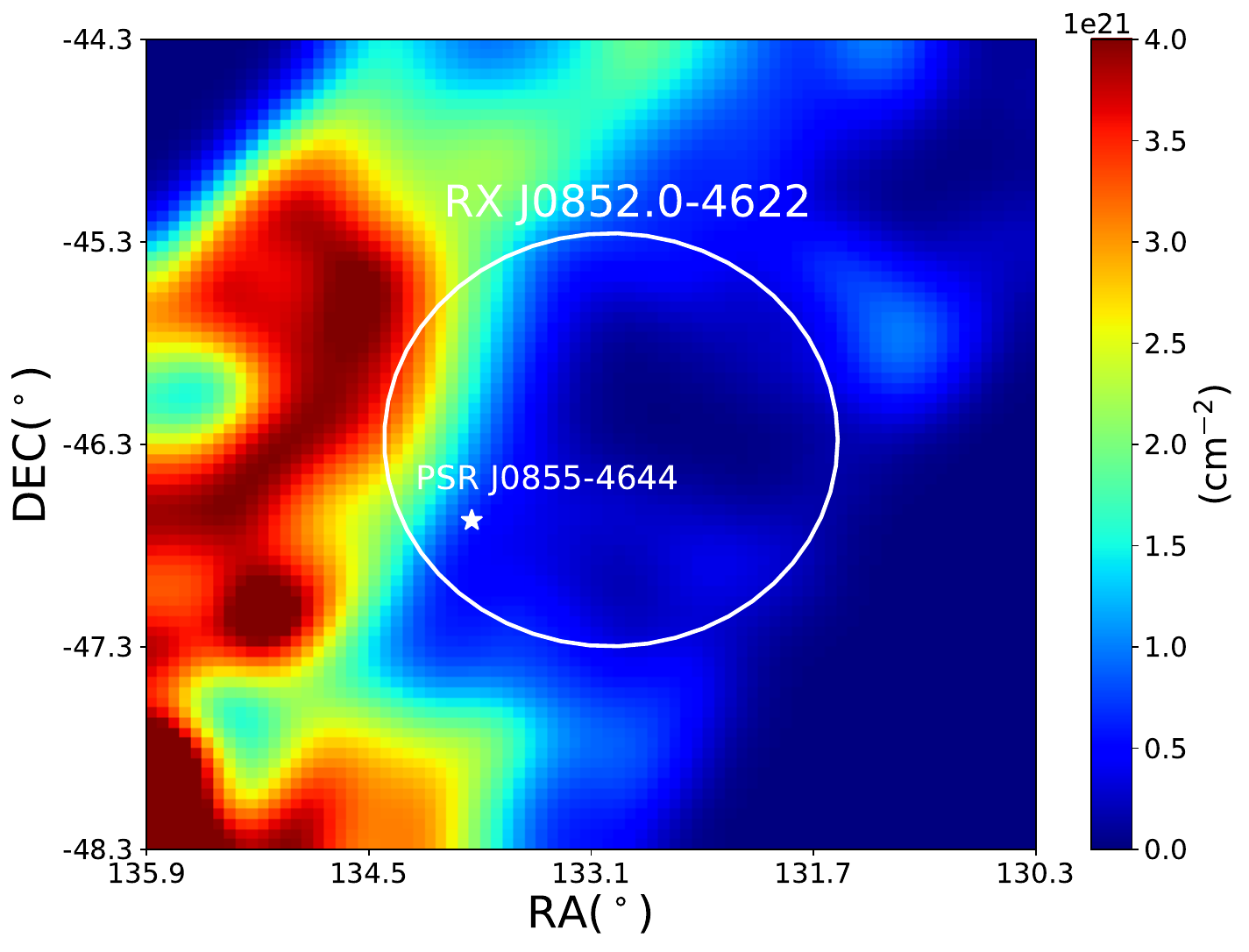}
\includegraphics[scale=0.3]{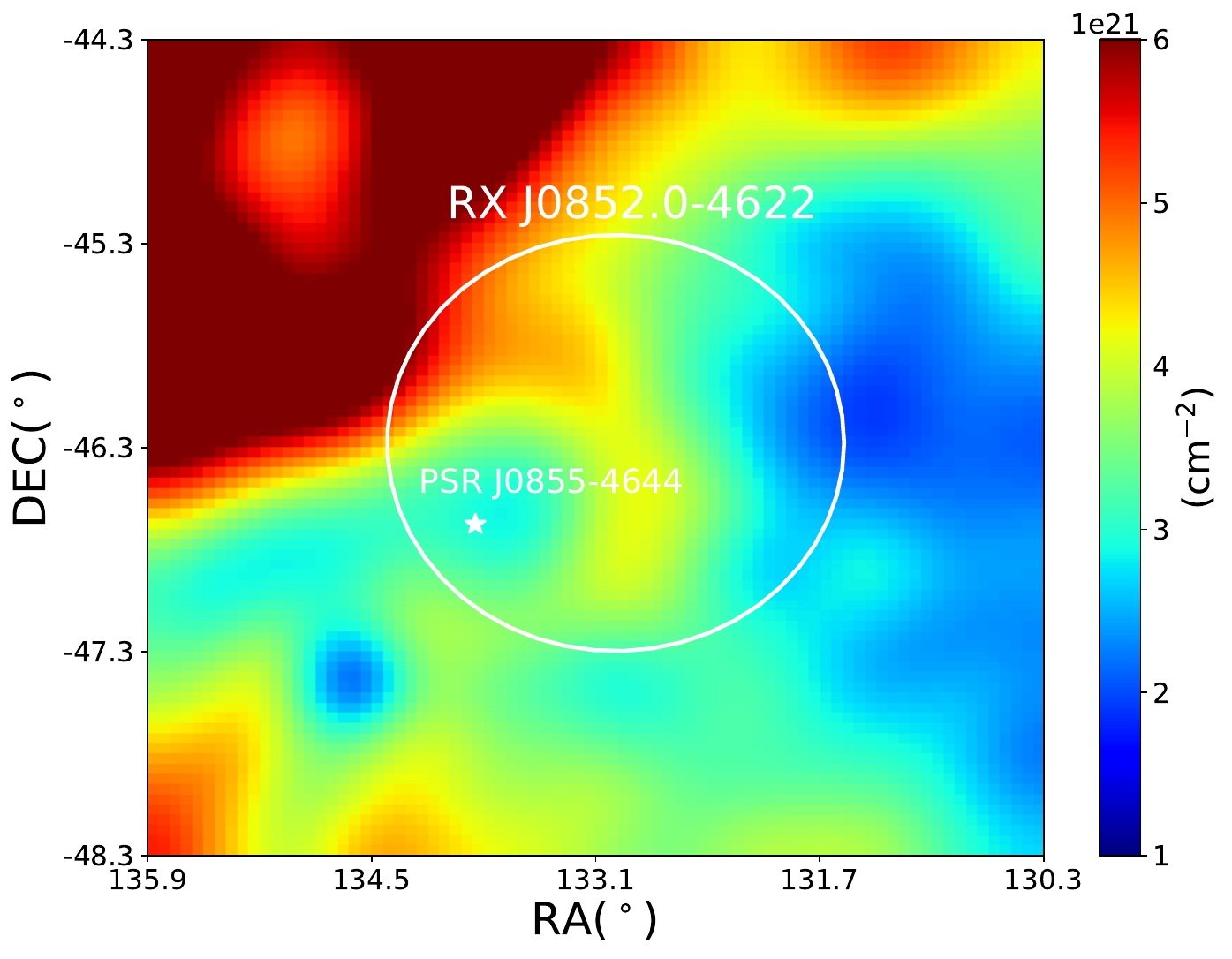}
\caption {Gas column densities (in units of $\rm cm^{-2}$) in different gas phases. The left panel shows the H$_{2}$ column density derived from CO data. The right panel shows the map of HI column density derived from the 21-cm all-sky survey \citep{2016HI4PI}. We integrate the gases within the velocity interval from $\rm -4\ km\ s^{-1}$ to $\rm 50\ km\ s^{-1}$. The white ring shows the position of the SNR RX J0852.0-4622, and the white star marks the position of the pulsar PSR J0855-4644.}
\label{fig:Gas}
\end{figure*}

For molecular gas, we use CO data from the CfA 1.2m millimetre-wave telescope to study the molecular cloud components in this region \citep{2001Dame}. The standard assumption of a linear relationship between the velocity-integrated brightness temperature of the CO 2.6-mm line, $W_{\rm CO}$, and the column density of molecular hydrogen, N($\rm H_{2}$), is expressed as N($\rm H_{2}$) = $X_{\rm CO} \times W_{\rm CO}$ \citep{1983Lebrun}.
Here, $X_{\rm CO}$ is the H$_{2}$ / CO conversion factor, which was chosen to be $\rm 2.0 \times 10^{20}\ cm^{-2}\ K^{-1}\ km^{-1}\ s$ as suggested by \citet{2001Dame} and \citet{2013Bolatto}. 
The velocity distribution of $\rm ^{12}CO$ (J=1-0) content shows excess in the velocity range of $\rm -4\ \ km\ s^{-1}$ to $\rm 50\ \ km\ s^{-1}$, which is suggested to be likely associated with the SNR \citep{2017Fukui}. 
The molecular gas column density map, as depicted in the left panel of Fig.~\ref{fig:Gas}, aligns with the findings reported by \citet{2007Aharonian}, indicating a similar distribution of molecular gas in the region.
In the eastern part of the remnant, due to the presence of the Vela molecular ridge, a high density region is clearly visible.

RX J0852.0-4622 is also located within a massive HI complex \citep{2017Fukui}. We use 21 cm HI emission line to trace neutral atomic gas, utilizing data from the data-cube of the HI $\rm{4\pi}$ survey (HI4PI) \citep{2016HI4PI}. 
The HI column density is calculated using the equation,
\begin{equation}
N_{HI} = -1.83 \times 10^{18}T_{\rm s}\int \mathrm{d}v\ {\rm ln} \left(1-\frac{T_{\rm B}}{T_{\rm s}-T_{\rm bg}}\right),
\end{equation}
where $T_{\rm bg} \approx 2.66\ \rm K$ is the brightness temperature of the cosmic microwave background radiation at 21 cm, and $T_{\rm B}$ is the brightness temperature of the HI emission. 
When $T_{\rm B} > T_{\rm s} - 5\ \rm K$, $T_{\rm B}$ is truncated to $T_{\rm s} - 5\ \rm K$, with $T_{s}$ set at 150 K. 
The HI column map for velocity from $\rm -4\ km\ s^{-1} $ to $\rm 50\ km\ s^{-1}$ is shown in the right panel of Fig.~\ref{fig:Gas}, which is consistent with the range in \citet{2017Fukui}. 
The HI column density shows spatial consistency with the centre of the SNR.
However, there is almost no ionized hydrogen ($\rm H_{II}$) gas in this region.

The total mass within each pixel of the cloud is determined using the following expression
\begin{equation}
M_{\rm H} = m_{\rm H} N_{\rm H} A_{\rm angular} d^{2}
\end{equation}
where $M_{\rm H}$ is the hydrogen mass, N$_{\rm H}$ = N$_{\rm HI}$ + 2N$_{\rm H_{2}}$ is the total hydrogen column density in each pixel, $A_{\rm angular}$ is the solid angle subtended by each pixel (as shown in Fig.~\ref{fig:Gas}), and $d$ is the distance from Earth to the SNR RX J0852.0-4622.
If we assume that the GeV $\gamma$-ray emission within the region has an angular size of $1.02^{\circ}$ for the whole SNR, we can calculate the total mass and number of hydrogen atoms in each pixel within the region. 
%If we assume that the GeV \gray\ emission within the region is spherical in geometry, with the corresponding size of $1.02\deg$  
The total mass in the GeV \gray\ emission region is estimated to be $\sim 1.33 \times 10^{5}~\msun$, as listed in Table~\ref{table:gas}. 
The total masses of the HI and the molecular gases are estimated to be $\sim  \rm{10^{5}\msun}$ and $\sim3.2\times \rm{10^{4}\msun}$, respectively, which is consistent with the measurements from \citet{2017Fukui}.
We then estimate the average density by distributing the velocity-integrated gas within a line-of-sight extended conical volume with a depth of $L=d=1.41$ kpc and the same projected angular size $1.02^{\circ}$.
The radius of the GeV \gray\ emission region is estimated as $r = d \times \theta (\rm rad) \sim 1.41\ \rm kpc \times (1.02\deg \times\ \pi / 180\deg) \sim 25\ \rm pc$, where $d$ is the distance from Earth to the objective region.
The total gas number density averaged over the volume of the GeV \gray\ emission region is n$_{\rm gas}$ = 5.8 cm$^{-3}$. 
Table~\ref{table:gas} provides details on the different gas masses and number densities within the GeV \gray\ emission region of SNR RX J0852.0-4622.

%----------------------------------------------------- FIGURE 4

% Don't change these lines
\bsp	% typesetting comment
\label{lastpage}
\end{document}